\documentclass[aps,10pt,prl,twocolumn,a4paper,footinbib,superscriptaddress,floatfix]{revtex4}

\usepackage{amsmath,amssymb}
\usepackage{hyperref}
\usepackage{bbm}
\usepackage{slashed}
\usepackage{graphicx}
\usepackage{mathptmx}
\usepackage{xcolor}
\usepackage{natbib}

\usepackage{verbatim}

\newcommand{\bw}{\begin{widetext}}
\newcommand{\ew}{\end{widetext}}

\newcommand{\be}{\begin{equation}}
\newcommand{\ee}{\end{equation}}
\newcommand{\bestar}{\begin{equation*}}
\newcommand{\eestar}{\end{equation*}}

\newcommand{\bi}{\begin{itemize}}
\newcommand{\ei}{\end{itemize}}
\newcommand{\bea}{\begin{eqnarray}}
\newcommand{\eea}{\end{eqnarray}}

\newcommand{\hbo}{\hbox to 1 true cm {\hfill } }

\newcommand{\ud}{\mathrm{d}}

\newcommand{\e}{\mathrm{e}}		

\begin{document}
\title{Pair annihilation in laser pulses: optical vs. XFEL regimes}
\author{Anton Ilderton}\email[]{anton.ilderton@physics.umu.se}
\author{Petter Johansson}
\author{Mattias Marklund}
\affiliation{Department of Physics, Ume\aa\ University, 901-87 Ume\aa, Sweden}

\begin{abstract}
\noindent  We discuss the theory and phenomenology of pair annihilation, within an ultra-short laser pulse, to a single photon. The signature of this process is the uni-directional emission of single photons with a fixed energy. We show that the cross section is significantly larger than for two-photon pair annihilation in vacuum, with XFEL parameters admitting a much clearer signal than optical beams. 
\end{abstract}
\maketitle

\paragraph{Introduction.} Prospects for probing the quantum vacuum using laser light at facilities such as ELI and the European XFEL currently attract a great deal of attention. Taking into account both the finite duration and spatial extent of modern laser pulses is challenging, but good analytic progress can be made if one assumes the laser is not too tightly focussed; it may then be modelled by a time dependent electric field, or a plane wave. Scattering in such backgrounds is very well understood, especially when the pulse is treated as having an effectively infinite duration \cite{Nikishov:1964zza, Nikishov:1964zz, Narozhnyi:1964}. These results have recently been re-evaluated, following the advent of ultra-short laser pulses,  in order to expose the effects of finite pulse geometry on laser-particle scattering \cite{Hebenstreit:2009km, Mackenroth:2010jk, Dumlu:2010vv, Mackenroth:2010jr, Dumlu:2011rr}.  One may therefore still ask if, within this well known topic, there are unique experimental signatures from which one could easily identify strong-field effects.

We discuss such a signal here, based on electron-positron annihilation. Unlike in vacuum, pair annihilation in a laser pulse may produce a single photon, both the energy and momentum of which are fixed. Photons from each annihilation event are emitted in the same direction and with the same energy.   In other words, repeated (idealised) events would themselves produce a laser-like beam of photons. Since the produced photons carry half the energy of the incoming, accelerated, particles, they have energies well above the electron rest mass: our annihilation process would produce a gamma ray beam! Even if gamma-ray lasers are still a pipe dream, we note that the most promising route to their developement is via the annihilation of molecular positronium, as has been demonstrated in the lab \cite{Cassidy}. The process we investigate here is essentially an `accelerator based' version of that idea. 

We begin by summarising the properties of our process, before giving its exact cross section in a laser pulse of finite duration, using strong-field QED. This is then investigated for a variety of realistic optical and XFEL beam parameters. For related recent investigations see \cite{Smolyansky:2010as, Hu:2011eq}.

\paragraph*{Kinematics.} In vacuum, momentum conservation implies that an electron-positron pair must annihilate to (at least) two photons \cite{DiracOrig}. In a laser field, though, the one-photon channel opens. This process is rather novel since there is only one allowed final state, modulo polarisations: momentum conservation alone determines the scattering products, as we now describe. We model the laser by a plane wave function of $k.x$, with the laser momentum $k_\mu$ lightlike. A pair with momenta $p_\mu$ and $p'_\mu$ (mass $m$) annihilate, using up some quantity $l \, k_\mu$ of momentum from the laser fields, and produce a single photon of momentum $k'_\mu$. Momentum conservation gives $p_\mu + p'_\mu + l k_\mu = k'_\mu$. There are four components to this equation, but a real photon has only three (momentum) degrees of freedom. This implies that the energy drawn from the laser is fixed: indeed, one finds after squaring up that $l$ is determined by the incoming momenta as $l = -(m^2+p.p')/k.(p+p')$. This constraint on $l$ gives the right counting of degrees of freedom. That $l$ is negative may be interpreted as energy being given up to the laser. Note that $l$ is not an integer, in general. (It effectively becomes so only for {\it periodic} backgrounds, after explicitly subtracting that part of $l$ which generates the electron mass shift \cite{Heinzl:2010vg,Review}.)  To illustrate, let the pulse move down the $z$ axis, so $k_\mu=m \nu (1,0,0,1)$ with $\nu= \omega/m$ the dimensionless laser frequency. The pair are introduced from the transverse directions with equal energy $E=m\gamma$ and opposite momenta. One easily finds from the above that $l=-\gamma/\nu$ and $k'_\mu = m\gamma(1,0,0,-1)$. Hence, half of the incoming energy is given up to the field and half (always) produces a backscattered photon with high frequency  $\omega'=m\gamma$. 

%
\begin{figure}[t!]
	\includegraphics[width=0.55\columnwidth]{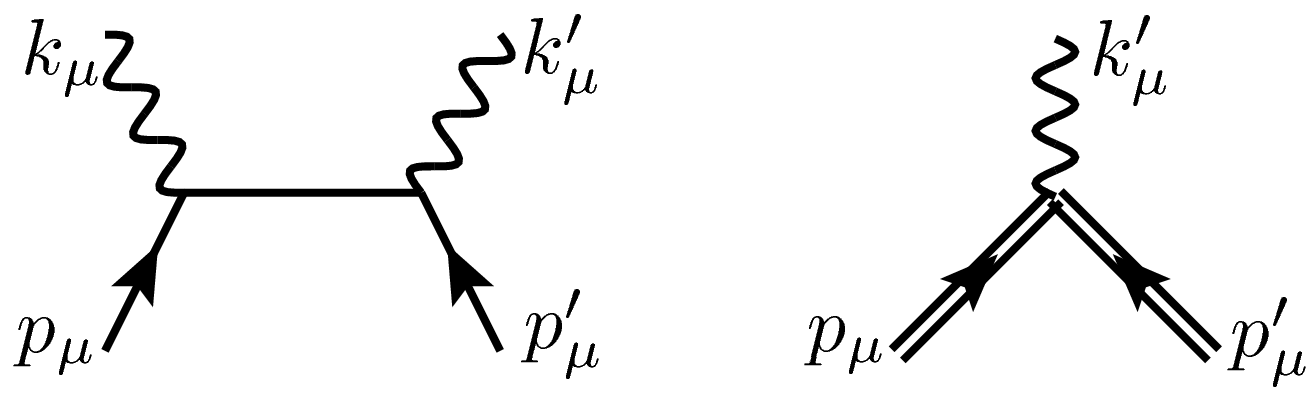}
	\caption{\label{Fig:Recomb} {\it Left}: annihilation to two photons in vacuum. Right: the one-photon channel in a laser pulse. Double lines indicate dressing of the fermions by the laser field.}
\end{figure}
\paragraph{The cross section.} The laser field strength is $F_{\mu\nu}(k.x) = \dot f_i(k.x) (k_\mu a^i_\nu - a^i_\mu k_\nu)$.  With $k_\mu$ as given above, the transverse polarisation vectors are $a^i_\mu = (ma_0/e)\delta_{i\mu}$ for $i=1,2$. Our normalisation is such that the invariant, dimensionless amplitude $a_0$ is the peak laser intensity.  The two functions $\dot f_1(k.x)$ and $\dot f_2(k.x)$ give the shape of the laser pulse. (The derivative on $f_j$ is for convenience, since it is the $f_j$ themselves which appear in the cross section.) Define $f_3 \equiv f_1^2 + f_2^2$. We take the fields to vanish outside of $k.x\in [0,2\pi N]$, so that $2\pi N$ is the {\it Lorentz invariant} pulse duration and $N$ the number of laser cycles. Introducing the pulse envelope function $g(k.x)$, such that $g$ vanishes smoothly at the pulse edges and $|g|\leq 1$, we parameterise $f_1(x)=g(x)\cos(x)$ and $f_2(x)=g(x)\sin(x)$.

In vacuum, a scattering cross section is given by the probability for a process to occur per unit volume per unit time (i.e.\ one divides out the interaction volume, which is the volume of spacetime, after squaring the $S$-matrix element), divided by the incoming particle flux. The cross section $\sigma$ for our process is defined analogously: the only difference is that the interaction volume is the pulse volume.  To proceed, we introduce functions of the external momenta:  
\bea
	\alpha_i = e\bigg( \frac{a_i.p}{k.p} - \frac{a_i.p'}{k.p'}\bigg) \;,\quad i  = \{1,2\}\;, 
\eea
along with $u=k.k'^2/(4k.pk.p')$ and $\alpha_3 = 2m^2a_0^2u/k.k'$, which appear in three `Bessel-like' integrals ($i=1,2,3$),
\be\label{BK}
	B_i = \int\limits_0^{2\pi N}\! \ud x\ f_i(x)\ \exp\bigg[ i l x-i\alpha_j\int\limits_0^x \!\ud\varphi\ f_j(\varphi)\bigg] \;.
\ee
Repeated indices are summed over $j\in \{1,2,3\}$. Define also $B_0 = (\alpha_j B_j)/l$ \cite{Ilderton:2010wr} and the combination
\bestar
	Z = 2|B_0|^2 + a_0^2(2u-1)(2 |B_1|^2 + 2 |B_2|^2-B_0 B_3^*-B_0^*B_3) \;.
\eestar
The flux is $I=\sqrt{(p.p')^2-m^4}$, and the cross section is
\be\label{SIGMA}
\sigma = \frac{1}{4I}\cdot\frac{e^2m^2}{2k.k'}\cdot\frac{1}{2\pi N}\cdot Z \;.
\ee
Some brief remarks on known cases. In a periodic laser field ($N=\infty$), (\ref{SIGMA}) is a delta comb \cite{Nikishov:1964zza, Nikishov:1964zz}: to make sense of this one must add various effects of finite duration by hand \cite{Review}. This problem does not appear at all if one includes the physical pulse duration from the outset, as we have done. In a periodic field, (\ref{SIGMA}) is a function of the effective electron mass, but this plays no explicit role here, as has been observed for other processes in short pulses \cite{Heinzl:2010vg,Mackenroth:2010jr}: the energy of the produced photon, for example, does not depend on the effective mass. 

In a constant (crossed) field, (\ref{SIGMA}) vanishes. To understand why, compare with another laser process in the constant field limit, say nonlinear Compton scattering \cite{Mackenroth:2010jr}. In that process, the interaction volume diverges, which appears to suppress the rate. However, it was shown in \cite{Nikishov:1964zza} that the integrals over the final states also contribute an infinite and equivalent factor, so the total cross section remains finite. On the other hand, the final state space in our process is trivial and independent of the details of the background, so no such cancellation can occur in general. Hence the emission rate vanishes.  We now examine $\sigma$ for various beam and particle parameters. 

\paragraph{Low intensity optical beams.}
We begin by following \cite{Voroshilo}, which considered an optical laser with intensity $a_0= 1\ldots 5$. This is outside the perturbative regime ($a_0<1$) but low by modern standards: ELI beams will have $a_0\sim 10^4$.  Without resorting to perturbation theory, \cite{Voroshilo} provided an elegant method for evaluating the functions (\ref{BK}). The integrals therein seem suited to evaluation by stationary phase, but the kinematics of the process forbid the exponents from having {\it any} stationary phase points. One may instead separate the integrand into slowly and rapidly varying parts  if the number of laser cycles is large, and then $N$ sets the rapid oscillation scale \cite{Seipt:2010ya}. (The method is particularly suited to the XFEL regime, as we will shortly see.)  This leads to an expansion of the integrand of (\ref{BK}) in Bessel functions $J_n$, similarly to the case of a periodic plane wave, for details see \cite[\S 101]{Berestetskii:1982}. The remaining rapidly oscillating exponent now admits stationary phase points, which are the solutions of
\begin{figure}[t!]
\centering\includegraphics[width=0.6\columnwidth]{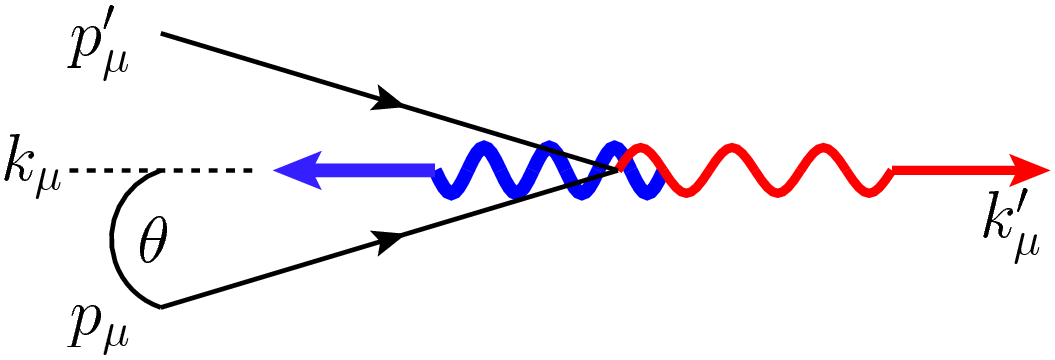}
\caption{\label{Fig:Voro} Collision geometry in the lab. The pair travel anti-parallel to the laser (blue, momentum $k_\mu$) with an offset angle $\theta$. The emitted photon (red, $k'_\mu$) is backscattered along the laser direction.}
\end{figure}
\be\label{StatPhase}
	g^2(x) = \frac{l-n}{\alpha_3} \;,\qquad n\in \mathbb{Z}.
\ee
For given $p_\mu$ and $p'_\mu$ (i.e.\ given $l$ and $\alpha_3$) there is a finite range of negative $n$ which contributes. For each $n$ the stationary phase points are calculated, and a sum over $n$ performed to obtain the $B_i$. (We have verified this approach using the powerful numerical methods of \cite{Moylan:2007fi}.) The rapid decay of the Bessel functions $J_n$ with increasing $n$ means that the most significant contributions to $\sigma$ come from low $n$. Since $g^2>0$ implies $l-n>0$, such contributions approximately require
\be\label{five}
	l\simeq -1 \implies 1\simeq \frac{\gamma}{\nu} (1-\beta\cos\theta) \;,
\ee
where $\beta=\sqrt{1-1/\gamma^2}$ as usual. The leading factor in (\ref{five}) increases with decreasing laser frequency, and so very small angles $\theta$ are required to keep the value of $l$ low.  An appreciable signal will therefore require the setup shown in Fig.~\ref{Fig:Voro}, with the pair's momenta a small angle $\theta$ from being anti-parallel to the laser direction \cite{TRANS}.  Assuming an optical frequency $\omega = 5$ eV ($\nu = 10^{-5}$), intensity $a_0\sim 1\ldots 5$ and a large number of cycles $N\gg1$ (this does not model a modern pulse, see below), along with high energy pairs, $\gamma\sim 10^6$, $\sigma$ exceeds that for the two-photon channel in vacuum by over an order of magnitude, for a range of collision angles $\theta$  \cite{Review, Voroshilo}.  This is where we encounter the difficulty in observation of our process: the required angle is $\theta^\circ\lesssim 10^{-4}$ because of the low optical frequency. Such precision seems experimentally challenging, at best, so we turn to alternative laser parameters. We are naturally led to consider a {\it high frequency} XFEL beam. We will see in this regime that not only does the cross section continue to dominate over that in vacuum, but that one gains several orders of magnitude in ``experimental ease".

\paragraph{The XFEL regime.} 
We assume a 100~fs pulse at X-ray frequency $\omega = \tfrac{1}{2}10^4$~eV ($\nu=1/100$), generated by the European XFEL at DESY \cite{XFEL,Ringwald:2001ib}, and $\gamma=80$ (already available at the DESY FZD linac). This pulse contains $N=1.2\times 10^5$ cycles due to its high frequency: the `slowly varying phase' approximation introduced above is therefore extremely well suited to the XFEL regime, as corrections go like $1/N$. To be concrete we choose the envelope function $g(x)\equiv \sin^4(x/2N)$ following \cite{Mackenroth:2010jk}. Our initial intensity is $a_0=1/10$, which will shortly be increased.

\begin{figure}[t!]
\centering\includegraphics[width=0.75\columnwidth]{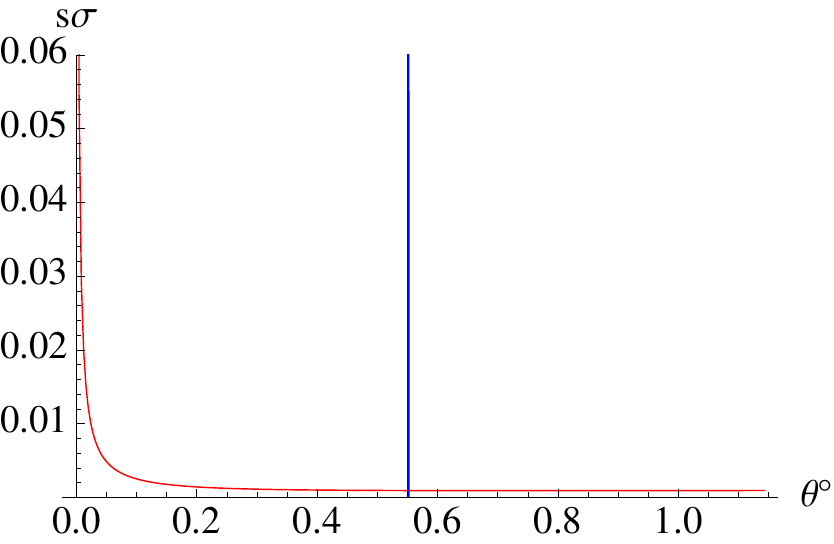}
\centering\includegraphics[width=0.75\columnwidth]{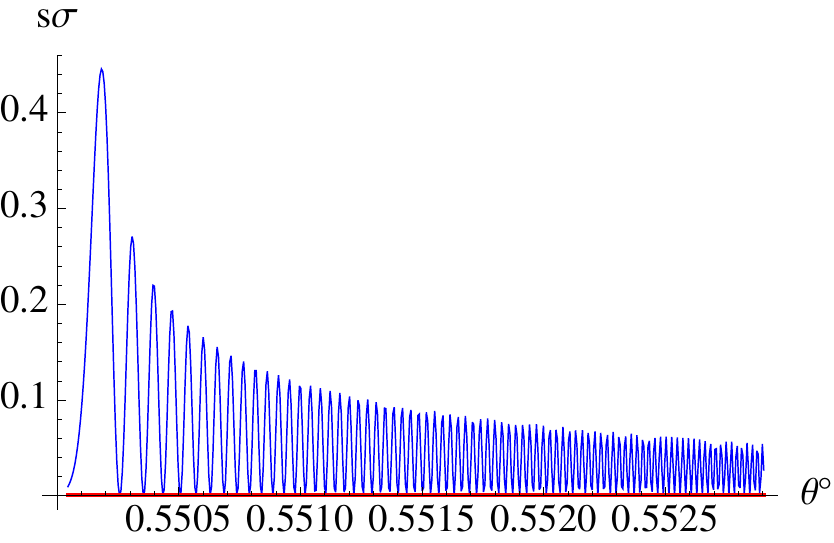}
\caption{\label{Fig:ZoomIn} Cross section $\sigma\times s$ ($\hbar=c=1$) in a 100 fs XFEL pulse of intensity $a_0=1/10$, $\nu=1/100$, with $\gamma=80$. {\it Top}: the cross section dominates over that in vacuum (red, bottom line) for a narrow angular range (vertical scale curtailed for easier comparison). {\it Bottom}: a closer view of the above peak. $\sigma$ exhibits a rich substructure.}
\end{figure}

Two views of the resulting cross section, multiplied by Mandelstam $s=(p+p')^2$, are shown in Fig.~\ref{Fig:ZoomIn} as a function of the collision angle $\theta$. Since our intensity is low we compare the cross section with that for two-photon annihilation in vacuum. There are clear quantitative and  qualitative differences: while the two photon process is supported over all $\theta$, the one-photon channel is very strongly peaked over a narrow angular range. In the lower panel of Fig.~\ref{Fig:ZoomIn} we `zoom in' on the cross section to expose its oscillatory substructure, which is an effect of finite pulse duration as discussed in \cite{Hebenstreit:2009km, Heinzl:2010vg, Mackenroth:2010jr}. While the support of the peak is now offset from $\theta=0$ by roughly half a degree, its angular range remains very narrow.

\begin{figure}[b!!!]
\centering\includegraphics[width=0.75\columnwidth]{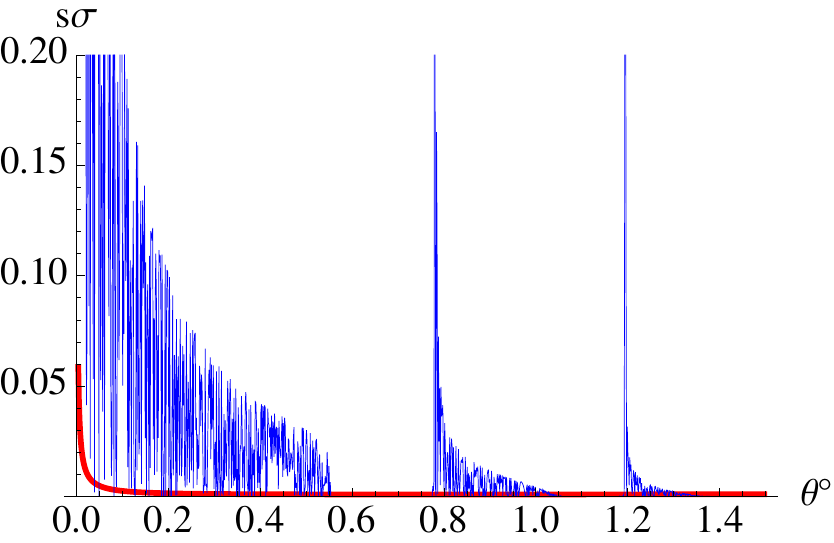}
\centering\includegraphics[width=0.75\columnwidth]{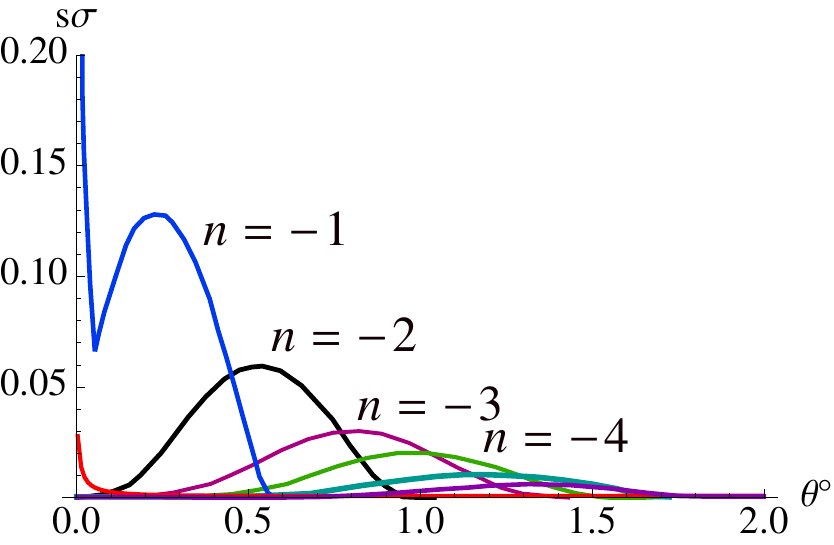}
\caption{\label{Fig:Sorted} At higher intensity the cross section is supported over a wider angular range of $\sim 2^\circ$. {\it Top}: at $a_0=1$ we again see a rapidly oscillating substructure. {\it Bottom}: at $a_0=10$ many $n$ must be summed over to obtain $\sigma$. Omitting the oscillating substructure, we have drawn the envelope functions of the first six, dominant, contributions.}
\end{figure}

The visibility of the signal is significantly improved by raising the XFEL intensity to the edge of the perturbative domain, $a_0=1$  (other parameters as above). The result is shown in the top panel of Fig.~\ref{Fig:Sorted}: the increased intensity leads to an appreciable signal over an angular range of $\sim1^\circ$. The appearance of disjoint signals (each with a rapidly oscillating substructure) corresponds to (\ref{StatPhase}) admitting solutions for $-n=1\ldots3$ over the given $\theta$ range: increasing the intensity allows more $n$ to contribute since the denominator in (\ref{StatPhase}) goes like $a_0^2$. The precise form of $\sigma$ is sensitive to the intensity, as illustrated in the lower panel of Fig.~\ref{Fig:Sorted} where we consider the `goal' XFEL intensity $a_0=10$, and observe a signal over $\sim 2^\circ$. Here, many $n$ give overlapping contributions to $\sigma$ \cite{Review}. Similar intensity-dependent behaviour is seen in nonlinear Compton scattering \cite{Harvey:2009ry}.  The two-photon result is shown in Fig.~\ref{Fig:Sorted} for comparison, but other processes may contribute in the nonperturbative domain $a_0\geq 1$, for example the two-photon channel in the laser (which is suppressed by a factor of $a_0^2$ perturbatively). The literature does not seem to contain a complete treatment of this process, though the case of a periodic field may be extracted by crossing symmetry from \cite{Hartin}, and prospects for coherent X-ray generation via this process are discussed in \cite{Henrich:2003uf}. 

Despite the challenge in detecting the signals presented here, we emphasise that the cross section is significantly greater than that in vacuum for the shown $\theta$ ranges. Furthermore, going to XFEL beams improves the visibility of the signal by four orders of magnitude: the angular range in Fig.~\ref{Fig:Sorted} is measured in whole degrees, $10^4$ times larger than that in the optical regime.

\paragraph{High intensity optical beams.}
The next generation of optical lasers will have extremely high intensities of $a_0\sim 10^4$, and ultrashort duration $N\sim\mathcal{O}(1)$. As such, neither a perturbative expansion in the intensity $a_0$ nor the above ``large $N$" method are applicable. Numerically, we have not found an appreciable signal for the geometry shown in Fig.~\ref{Fig:Voro} and reasonable collision angle. We therefore return to the transverse collision discussed in the introduction, which is a more natural experimental setup. We take $a_0=10^4$ and $\omega=1$ eV ($\nu = 2\times10^{-6}$), modelling an ELI strength optical laser. With this, each of the terms in the exponents of (\ref{BK}) is large:
\be\label{params}
	l \sim -10^6\gamma, \quad \alpha_{1,2} \sim - 10^{10}\beta\;, \quad \alpha_3 \sim 10^{14}/\gamma \;.
\ee
Assuming high energy incoming pairs, say $\gamma\sim 10^4$, each of the parameters in (\ref{params}) is around $\sim 10^{10}\equiv M$. The functions (\ref{BK}) in this regime therefore take the form
\be\label{BK2}
	B_i \sim \int\limits_0^{2\pi N}\! \ud x\ f_i(x)\ \e^{i M h(x)} \;,
\ee
Since the full exponent contains no stationary phase points ($h'\not=0$), one may generate an asymptotic expansion of these integrals using the standard `integration by parts' method:
\be\label{BK2}
	B_i \sim \frac{f_i(x)}{M h'(x)} \e^{i M h(x)}\bigg|_0^{2\pi N} +\ldots \;.
\ee
This is just the statement that the functions (\ref{BK}) are dominated, for short pulses and at high energy, high intensity, by edge effects. If the pulse turns on and off symmetrically then $f_i=0$ at the pulse edges, and the above term is zero. Going to higher orders in the expansion, one encounters the derivatives of $f_i$ at the pulse edges. At least the first derivative must vanish if we are to have a smooth pulse. The first nonvanishing term in the expansion (\ref{BK2}) therefore goes like $M^{-r-1}$, with $r$ the order of the first non-vanishing derivative of $f_i$ at the pulse edge: the smoother the pule,  the smaller the resulting cross section.

\paragraph{Relevance to cascade formation.}
It has been suggested that a cascade of particles could be triggered even by a single pair creation event in an intense laser pulse \cite{Fedotov:2010ja, Sokolov:2010am, Elkina:2010up}. 
Cascade codes use cross sections calculated in a constant, crossed field, as this is how any laser field looks {\it locally}, at sufficiently high intensity \cite{Nikishov:1964zza}. As discussed above, the cross section (\ref{SIGMA}) in such a field vanishes \cite{Nikishov:1964zza,Nikishov:1964zz}, and so it seems that one-photon annihilation is not particularly relevant to cascades. Even with finite pulse duration, we have seen that $\sigma$ is appreciable only for particular collision geometries with small angular tolerances, and such fine tuning is of course impossible in a cascade event. (Density effects \cite{Plasma} and higher order processes such as the two-photon channel \cite{Hartin} are a different matter, and must be discussed elsewhere.)

\paragraph{Conclusions.} We have discussed pair annihilation to one photon, within a laser pulse. This process has a trivial final state space and cannot occur in vacuum. We have provided a modern treatment which incorporates, from the outset, physical effects due to finite pulse duration. This removes the theoretical ambiguities which appear for periodic fields. We have analysed the cross section for beam parameters which are, or will soon be, realisable at the European XFEL and ELI.

We have seen that the cross section for our process can be significantly larger than that of the two-photon channel in vacuum, but that its observation requires fine tuning of the collision geometry. The best situation we have found is for a reasonably intense XFEL beam, in which the angular precision required is measured in degrees. This is an improvement from {\it thousandths} of degrees in the optical regime and is due to the high XFEL frequency. After submitting this paper the preprint \cite{Blaschke:2011is} appeared which supports these conclusions using a quantum kinetic approach.

A.~I.\ thanks T.~Heinzl and H.~Ruhl for very useful discussions. This work is supported by the European Research Council, Contract 204059-QPQV, and the Swedish Research Council, Contract 2007-4422.


\begin{thebibliography}{99}
\bibitem{TRANS} The pair's momenta lie in the same transverse plane. Dependence on this choice of azimuthal angle is weak and not discussed here.
\end{thebibliography}
\end{document}